\documentclass[conference]{IEEEtran}
\usepackage{graphicx}

\ifCLASSINFOpdf
\else
\fi
\usepackage[ruled]{algorithm2e}

\begin{document}
%
\title{Power Efficient Resource Allocation for Clouds Using Ant Colony Framework}

\author{\IEEEauthorblockN{Lskrao Chimakurthi}
\IEEEauthorblockA{Department of Computer Science and Engineering\\
National Institute of Technology\\
Calicut, Kerala 673601.\\
Email: siva.chimakurthi@gmail.com}
\and
\IEEEauthorblockN{Madhu Kumar S D}
\IEEEauthorblockA{Department of Computer Science and Engineering\\
National Institute of Technology\\
Calicut, Kerala 673601.\\
Email: madhu@nitc.ac.in}
}


%

\maketitle               

\begin{abstract} 
Cloud computing is one of the rapidly improving technologies. It provides scalable resources needed for the applications hosted on it. 
As cloud-based services become more dynamic, resource provisioning becomes more challenging. The QoS constrained resource allocation 
problem is considered in this paper, in which customers are willing to host their applications on the provider's cloud 
with a given SLA requirements for performance such as throughput and response time. Since, the data centers hosting the 
applications consume huge amounts of energy and cause huge operational costs, solutions that reduce energy 
consumption as well as operational costs are gaining importance. In this work, we propose an energy efficient mechanism 
that allocates the cloud resources to the applications without violating the given service level agreements(SLA) using Ant colony framework. 
\end{abstract}


\section{Introduction}

Cloud computing provides the utility services based on the pay-as-you go model. Users can host different kinds 
of applications on the cloud ranging from simple web applications to scientific workloads. These applications are delivered as services over 
the internet. Cloud customers no longer need to worry about the costs associated with under-provisioning and over provisioining. Many researchers focused mainly 
on hosting high performance applications in clouds without considering the energy efficiency. Since the energy costs are increasing, 
the need for optimising cost of data center resources is also increasing. This will not only reduce the energy consumption but also 
the operational cost. So, cloud resources need to be allocated not only to satisfy QoS requirements specified by users 
through SLAs, but also to reduce energy usage.\\ \\
The load on the servers in data center changes dynamically. So, we need efficient mechanisms to take this dynamism into 
account and allocate the resources to the services so that minimum number of servers will be used for hosting the services. 
Thus, we can achieve less operational cost and energy consumption. As the Ant colony\cite{14} mechanisms 
are helpful for adapting to dynamic behaviour of the loads, we believe that the above objectives can be achieved using 
intelligent ant agents for monitoring.

In this paper, we present a mechanism for adaptive resource allocation in cloud computing environments for hosting the 
applications with given QoS requirements as throughput and response time that reduces the power consumption of data center resources 
by considering the dynamic loads of servers using different ant agents.\\ \\
The rest of the paper is organised as follows. In the next section, we present some of the related work in this direction. 
Section III describes the system architecture used for the resource allocation. In section IV, the details about 
our methodology and the algorithms are presented. Section V outlines the future work and section VI concludes the paper.

\section{Related Work}

Resource allocation for clouds has been studied very extensively in the literature. The problem of deciding on an optimal 
assignment of requests to resources allocator is NP-hard\cite{4}. Several heuristic algorithms have been proposed 
by researchers for optimal allocation of cloud resources.

Market-based resource management\cite{6} has been proposed by Rajkumar Buyya et al. to manage the allocations of computing
resources. The current state-of-the-art in cloud computing had its limits in considering energy awareness. 
In \cite{2}, challenges and architectural elements for energy-efficient management of cloud 
computing environments are discussed. Also, they developed algorithms for dynamic resource provisioning and 
allocations that will improve the data center energy efficiency.

An energy aware resource allocation method for clouds has been proposed in \cite{8} that maximises the service provider's revenues. However, 
it is developed mainly for scheduling the servers for completing the customers jobs and not applicable for scheduling different applications that need critical performance requirements.
 Some other techniques also have been proposed for predicting the workloads of the servers by applying statistical methods\cite{15}.

The energy-efficient resource allocation solutions proposed in \cite{9} are not applicable for cloud computing because 
their only focus is on minimising energy consumption without considering dynamic service requirements of customers.

Ant algorithms are one of the most popular examples of $swarm$ $intelligence$ systems, in which a number of ant-inspired 
agents which are specialised in particular sophisticated functionality follow simple rules with no centralized control.
 The complex global behavior emerge from their local interactions using pheromone\cite{1}. Bio-inspired techniques have already 
been applied to solve a number of complex problems, such as task allocation, routing, graph partitioning, etc. 
Ant colony mechanisms are more applicable to grid scheduling\cite{3}.

An ant colony framework was proposed in \cite{4} for clouds that can be applied to scheduling of online jobs and batch jobs. 
The work in \cite{5} is concerned with self organising techniques for deployment of virtual machine images onto physical machines, which reside in different 
parts of the network, have been proposed in order to improve the scalability and to handle the dynamism using Cross-Entropy 
Ant system. It tries to find out the optimal deployment mappings of virtual machines to the nodes in the cloud. 
\section{Our System Architecture}
The main aim of our resource allocation is to allocate the online service requests for applications which are CPU and memory intensive.
To achieve the objective of adaptive resource allocation for satisfying the service requests 
of customers, we use the following architecture.\\
\begin{figure}[h]
 \centering
 \includegraphics[scale=.35]{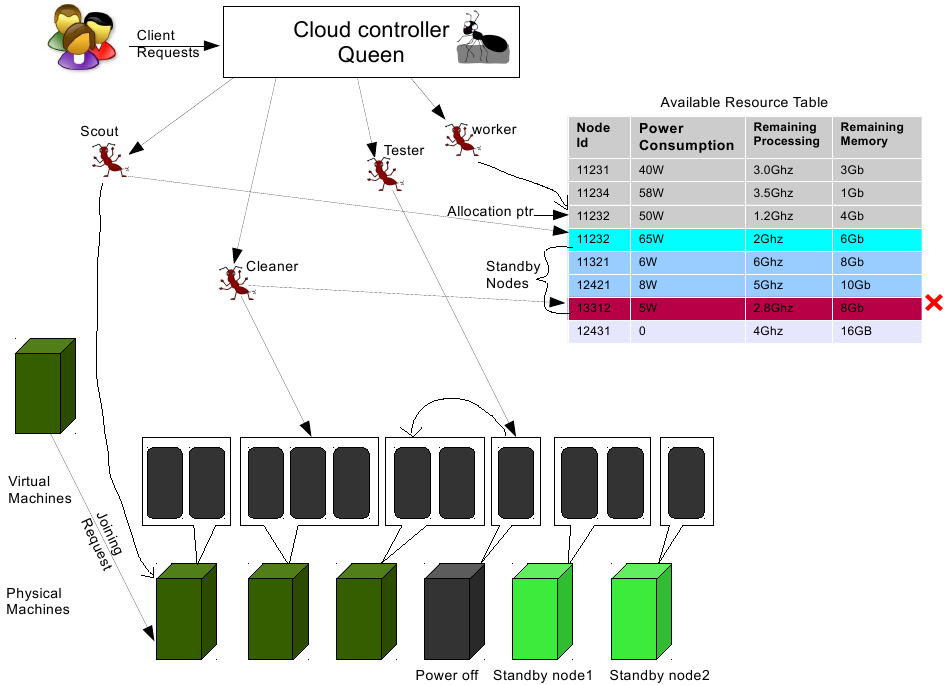}
 \caption{Cloud Architecture with Ant Agents }
 \label{fig:1}
\end{figure}

\textit{Users/Brokers:} Users or brokers acting on their behalf submit service requests to the cloud via cloud controller for processing.

\textit{Cloud Controller:} It acts as the interface between the cloud service provider and external users/brokers. It 
acts similar to the Queen in the ant colony.

\textit{Virtual Machines(VMs):} This is where the applications of customers will be deployed. We can dynamically create, start, stop and migrate 
these VMs depending on our requirement, from one physical machine to another.

\textit{Physical Machines:} These are the physical computing servers that will provide hardware infrastructure for creating virtual machines.

\section{Our Methodology}
We gather the power consumption of each server in the data center along with the resource capabilities such as CPU processing 
power and primary memory before admitting them into cloud. We store this information consisting of \textit{Node Id, Processing Power, Memory and Power Comsumption}
 in a table.\\ \\
We sort the nodes in the descending order of the following metrics.\\

$\frac{Processing \hspace{1 mm} power \hspace{1 mm} of \hspace{1 mm} the \hspace{1 mm} node \hspace{1 mm} (in \hspace{1 mm} Ghz)}
{Power \hspace{1 mm} consumption \hspace{1 mm} of \hspace{1 mm} CPU \hspace{1 mm} (in \hspace{1 mm} Watts)} = {PPW}$
\\ 

$\frac{Memory \hspace{1 mm} capacity \hspace{1 mm} of \hspace{1 mm} node \hspace{1 mm} (in \hspace{1 mm} Gb)}{Power \hspace{1 mm} consumption \hspace{1 mm} of \hspace{1 mm} Memory \hspace{1 mm} (in \hspace{1 mm} Watts)} = {MPW}$ \\

\textit{PPW = Processing Power per Watt, MPW = Memory Consumption per Watt}\\ \\
We take the power consumption as the power consumed by CPU or memory when their utilisation is 100\% which is measured before admitting the node in the cloud.
We assume that the power consumed by all the remaining components of a node are same for all the nodes. We consider that, all the nodes 
are having the same network connectivity and access to the shared persistent storage space that stores the VM disk images. \\ \\
We store this information in a table called \textit{Available Resource Table} with \textit{Available node's Id}, current power consumption and remaining capacity(updated by the gathered information from ant agents) as shown in fig 1.
This table is represented as an \textit{Array List} with a pointer \textit{(Allocation ptr)} being pointed to the node on which next service request is deployed. This pointer will be adjusted by different ant agents.\\ \\
 We followed the terminology related to different types of ant agents used in \cite{3}. The functionality of these agents is explained in the subsections below.
\subsection{\textbf{Cloud Controller \& Queen Ant:}}
The requests from the customers consisting of the following, are given to the controller.\\ \\
(i) Throughput(THPUT) (In \%)\\
(ii) Avg. Response Time(RTIME)\\
(iii) Application Code\\
(iv) Operating System\\ \\
Cloud controller maintains a queue(Q) for storing the service requests for hosting the applications. It enqueues each of the service request received, in this queue.\\ \\
It generates the tester, scout, cleaner and worker ants periodically. The movement of these ant agents is modelled in the following way.

  Each ant except Queen \& Worker maintains a \textit{Visited Node list} which is initially empty. Each node in the cloud maintains a list of 
neighbouring node's information. Whenever an ant reaches a node, it updates the controller about the current utilisation and 
randomly chooses an unvisited neighbouring node. When all the nodes are covered, it makes the \textit{Visited Node list} empty and continues again in the 
same way.\\ \\
We can change the number of ants that will be produced so that it will yield better results depending on our requirement.
The next subsection describes the method used by worker ants for accepting or rejecting the service requests. 
\subsection{\textbf{Worker Ant:}}
Whenever a service request received in the queue, one of the worker ants creates a VM with a specific CPU processing power and memory etc, if accepted. 
So, worker ants are always looking in the queue to check if there are some pending requests to be processed. If such a request is found, it dequeues the request and calls $Algorithm$ $1$.\\ \\
\begin{algorithm}
\KwIn{Request for hosting a client application.}
\KwOut{Accept and host the application on a VM (or) Reject request.}
\Begin{
\If{ArrayList pointer currently points to a valid node}
{
\If{The pointed node has enough resources for creating a basic VM}
{
Create a VM with the requested OS and deploy the application on it;\\
Create SLA monitor agent to monitor it;\\
\If{Remaining capacity of node doesn't have min. resources for hosting a “basic” VM}
{
Move the ArrayList pointer to the next node in the list;\\
\If{ArrayList pointer points to invalid node}{Notify\_Admin(Resource Scarcity);}
}
}
\Else{Move the pointer to the next node in the list;\\
\If{The node is capable of hosting a basic VM}{Call Allocate(Q);}
}
}
\Else{Reject the request;}
}
\caption{Allocation of client requests by Worker ant}
\end{algorithm}
Since most of the CPUs are work conserving, we are creating a VM like Amazon Standard Instance\cite{10} with specific CPU processing power and memory. Depending on the load, more intensive applications can use the resources of the other VMs having less load.\\ \\
The worker ant is only responsible for deploying the request on a VM as shown in $Algorithm$ $1$. Load balancing decisions are taken by tester ant. 
After deploying, it creates a Service Level Agreement(SLA) monitor agent that monitors the hosted application. In the next subsection, we provide the 
details about the SLA monitor agent.
\subsection{\textbf{SLA Monitor Agent:}}
It calculates the \textit{Avg. response time} and \textit{throughput} of the hosted application by continuously monitoring it. 
It passes this information to the hypervisor on that host in the form of a variable(SLAM) which is calculated depending on the performance of the application as shown in $Algorithm$ $2$. 
When the tester ant queries the node for utilisation information, hypervisor will send this SLAM value along with utilisation information. \\ \\
0. No need for balancing\\
1. Recommended for balancing

11. Migrate $\quad$12. Clone\\
2. Strictly needed (Critical)

21. Migrate $\quad$22. Clone\\ \\
If the SLAM value is 1, the tester ant will try to allocate this VM to a better node that have more available resources than 
current node among the currently running nodes and will not try to wake up a new node.
If it's value is 2 then, it will wake up the next power efficient standby node if needed as it is required to handle the heavier loads.
Depending on these values, the tester ant balances the load.

\begin{algorithm}
\KwIn{Throughput and Response Time}
\KwOut{SLAM value}
\Begin{
\If{ Response time $\textless$ 0.9$\times$RTIME AND Throughput $\textgreater$ 1.1$\times$THPUT }{return 0;}
\ElseIf{0.9$\times$RTIME $\textless$ Response time $\textless$ 0.95$\times$RTIME \\(OR) 1.10$\times$THPUT $\textgreater$ Throughput $\textgreater$ 1.05$\times$THPUT}
{
\If{Both cond1 \& cond2 are true}{return 12;}
\If{Only cond1 is true}{return 12;}
\If{Only cond2 is true}{return 11;}
}
\ElseIf{Response time $\textgreater$ 0.95$\times$RTIME \\(OR) Throughput $\textless$ 1.05$\times$THPUT}
{
\If{Both cond1 \& cond2 are true}{return 22;}
\If{Only cond1 is true}{return 22;}
\If{Only cond2 is true}{return 21;}
}
}
\caption{Behaviour of SLA monitor Agent}
\end{algorithm}
When the response time is 10\% less than given SLA time and 10\% greater throughput, then we will consider it as nomal situation. 
When the response time increases and is still 5-10\% less than given or 5-10\% more than throughput, then one of the tester ants will balance the load if some 
currently running nodes are lightly loaded. If the response time is reaching the SLA time and is just below 5\% or throughput is 5\% greater, 
then one of the tester ants will defintely move the VM to another node or clone this VM on another node and configure the router so that the traffic will 
be shared between the two. All the above values are tunable parameters and we can adjust them dynamically according to the requirement.

In the next subsection, we present the algorithms used by tester ant for taking the load balancing decisions based on the information received from hypervisor.
\subsection{\textbf{Tester Ant:}} 
The main job of the tester ants is to get the utilisation and power consumption information from each of the node and to update the 
available node's list. It also takes the load balancing decisions.  \\ \\
$\textit{Assumption 1:}$ Methods are available that will be able to estimate the fraction of memory actively used by each VM on a 
host like Memory Sampling used by Vmware ESX servers\cite{12}.\\ \\
$\textit{Assumption 2:}$ We are not not giving any preferential shares to the applications running on the same host.
We can use the idle memory taxation technique used by ESX servers to reclaim the unused memory from lightly loaded VMs and reallocate this 
to the memory contending VMs\cite{12}. Since most of the hypervisors are of work-conserving type, CPU processing power is used 
relative to the demand on the VMs instead of strict CPU cycle shares.\\

Based on above assumptions, we consider the load balancing on a node will be taken care by the hypervisor. 
We provide the algorithms for load balancing of various hosts in the cloud.

We consider that the utilisation of 80\% of CPU and 80\% of memory of a node is considered to be desirable utilisation and above 90\% 
is considered to be peak. However, we can change them according to our requirement.\\ \\
We get the utilisation of CPU and memory information by probing the $proc$ file system in Linux and resource utilisation in Windows. 
The tester ant will update this information in the $Available$ $node's$ $list$ along with the current power consumption shown by $powertop$ utility\cite{16}.
We have used the existing utilities for measuring power consumption because the existing power models, based on the utilisation, 
are not accurate at measuring it\cite{11}.\\ \\
For each of the visited node in cloud, the actions performed by the tester ants are coded in Algorithm 3.
\begin{algorithm}
\KwIn{CPU utilisation, Memory utilisation and SLAM value}
\KwOut{Load balanced node (or) Changing of a node to standby node}
\Begin{
\If{CPU (or) Memory utilisation $\textgreater$ 90\%}
{
Sort the VMs in the decending order of their utilisation and select the first;\\
\If{There is a VMs whose SLAM value = 2}{Call Algorithm 3;}
\If{There is a VMs whose SLAM value = 1}{Call Algorithm 4;}
}
\If {CPU and Memory Utilisation $\textless$ 50\%}
{
Get the Remaining Capacities of all the nodes above this node in Available Node's List;\\
Sort them in decending order of remaining capacity;\\
Sort the VMs on this node in the decending order of current Utilisation;\\
\If {It is possible to migrate all the VMs in this node to some nodes above in the list}
{
Migrate all the VMs;\\
Put this node in Standby mode;\\
Turn off the Standby node which is currently last in Available node's list;
}
}
}
\caption{Behaviour of Tester Ant}
\end{algorithm}

\begin{algorithm}
 \KwIn{CPU utilisation, Memory utilisation and SLAM value}
\KwOut{Load balanced node}
\Begin{
\If {There are VMs whose SLAM value = 22}
{Search for nodes in the Available node’s list that have remaining resources as 50\% of
current utilisation of the critical VM;\\
\If{Some nodes are available having enough resources}
{Select the first power efficient node;\\
\If{Selected node is Standby node}
{Wake-Up the selected node;\\
Turn On next node in the Available node list and put it in Stand by mode;\\
\If {No next node present}{Notify\_Admin (Few Resources);}
}
Create a clone of VM with resource entitlement as 50\% of current utilisation;\\
Configure the router such that the requests will be shared between these two VMs.\\
}
\Else{Notify\_Admin(Resource Scarcity);}
}
\If {There are VMs whose SLA value = 21}
{
Search for a nodes in the Available node’s list that are under utilised;\\
\If {Some nodes have free resources 30\% more than current utilisation of the critical VM}
{
Select the first power efficient node;\\
\If {Selected node is Standby node}
{Wake-Up the selected node;\\
Turn On next node in the Available node list and put it in stand by mode;\\
\If {No next node present}{ Notify\_Admin(Few Resources);}
}
Migrate the above VM to the selected node;\\
}
\ElseIf {some node is available and has only resources 50\% of current utilisation}
{
Place the cloned critical VM with resources as 50\% of current utilisation on this node;\\
Associate a flag with this VM so that cleaner ants will distinguish this from normal VMs;\\
Configure the router such that the traffic will be shared between the two VMs;\\
}
\Else{Notify\_Admin(Resource Scarcity);}
}
}
\caption{}
\end{algorithm}

\begin{algorithm}
 \KwIn{CPU utilisation, Memory utilisation and SLAM value}
\KwOut{Load balanced node}
\Begin{
\If{There are VMs whose SLA value = 11}
{
Sort these VMs in the descending order of utilisation and select the first; \\
Start searching from the first node in the list till the node which is above the current node in the list;\\
\If{Some nodes have 30\% more resources than current resource consumption of VM}{Migrate this VM to first power efficient node;} 
}

\If{There are VMs whose SLA value = 12}
{
Start looking from the first node in the list till the node
which is above the current node in the list;\\
\If {An available node has remaining resources of atleast 50\% consumption of current VM}
{
Place the clone of VM with resources as 50\% of the current utilisation on this node;\\
Associate a flag with this VM so that cleaner ants will distinguish it from normal VMs;\\
Configure the router such that the traffic will be shared between the two VMs;\\
}
}
\caption{}
}
\end{algorithm}

In order to improve the process of creating the VMs, we put three nodes below the current allocation node in standby mode and create the 
VMs with specific operating systems to be able to configure the applications on them quickly.
We prefer standby mode to hibernate mode because it requires less time to wakeup a node from standby mode 
than from hibernation. We consider that the overhead of VM migration is negligible.

The next subsection describes about the node discovery and registration which will be done by scout ants.
\subsection{\textbf{Scout Ant:}}
The aim of scout ant is to discover the newly added cloud nodes providing computing and memory services. When such a new node is found, 
it adds it to the available resource table. It is done through node registration.\\ \\
\textit{Node Registration:}

The node which wants to join the cloud must have to inform one of the nodes in the cloud by sending a request. 
When the scout ant visits a node and finds a request for joining, it registers the node with a unique id. 
It updates this information in the available node's list with resource utilisation and power consumption and places this node 
in appropriate position in the list.
Whenever a node is added by administrator in the event of resource scarcity, then the registration will be done by cloud controller node.\\ \\
$\textit{Assumption 3:}$ We assume that the node that is completing the registration for newly joining node will not get failed during the registration 
process. If it fails before registration, new node will contact another node in cloud.
\subsection{\textbf{Cleaner Ant:}}
It maintains the available resource table by removing the unavailable resources from the list. When this agent reaches a 
node in cloud and if it didn't respond to that agent for a specific time duration, then it assumes that this node is failed and it takes 
necessary actions for recovery and will remove this node information from the available node's list.

It removes the cloned VMs from the nodes if they are under utilised or if the performance is more than required. It also removes the VMs from the nodes whose service agreement get expired. It sends an alert to the customer before some days so that 
the service can be renewed if they needed.\\ \\
For each of the visited node in cloud, the actions performed by the cleaner ants are coded in $Algorithm$ $6$.
\begin{algorithm}
\KwIn{CPU and Memory utilisation}
\KwOut{Cloned VM removed if underlised and Notify user and admin when service time about to complete}
\Begin{
Look for the cloned VMs.\\
\If{The current resource consumption is NIL and its parent VM's utilisation $\textless$ 90\%}
{Remove this VM from this host;}
\If{Response Time and Throughput are 30\% more than that of SLA Values and parent is not utilisaing at its peak load}
{Remove this VM from this host;}

\If{A VM lease time remaining is less than a week} {Notify\_User(Service Time is about to complete);}
\If{A VM service time is over}{Remove VM from the node;}
}
\caption{Behaviour of Cleaner Ant}
\end{algorithm}

\section{Future directions}
We have investigated several cloud computing testbeds and found that cloudsim simulator$\cite{13}$ is 
suitable for testing the proposed mechanism. So, we are in the process of implementing the proposed mechanism on the cloudsim toolkit and the performance evaluation is in progress. 
We also plan to improve this by incorporating the load prediction and usage models so that this can be applied to real cloud environments.
\section{Conclusion}
We have proposed a power efficient, agent based solution for allocation of resources to cloud applications. We believe that this mechanism 
is very flexible and can be extended with improvements, as the solution modules are modelled as independent intelligent agents. We can incorporate 
additional functionalities in any of these ant agents.

\bibliographystyle{IEEEtranS}
\nocite{*}
\bibliography{bare_conf}		


\end{document}